\documentclass[10pt,aps,prfluids,print,longbibliography,superscriptaddress,notitlepage]{revtex4-2}
\usepackage{amsmath,amsfonts,bm, color,amssymb}
\usepackage{graphicx,overpic}
\usepackage{epstopdf}
\usepackage{mathtools}
\usepackage{url}
\usepackage{bm}
\usepackage{multirow}
\usepackage{soul}
\usepackage[%
  colorlinks=true,
  urlcolor=blue,
  linkcolor=blue,
  citecolor=blue
]{hyperref}

\graphicspath{{/}}%experiment/IMG
\usepackage{color}

\newcommand{\sigm}{{\sigma}}
\newcommand{\eps}{{\varepsilon}}

\input epsf

\newcommand{\Lr}{\mbox{$\lambda$}}

\newcommand{\out}{{\mathrm{}}}
\newcommand{\ins}{{d}}

\newcommand{\visrat}{{\lambda}}

\newcommand{\zhat}{{\bf \hat z}}

\newcommand{\that}{{\bf \hat \theta}}

\providecommand\bnabla{\boldsymbol{\nabla}}
\providecommand\bcdot{\boldsymbol{\cdot}}

\def\Xint#1{\mathchoice
   {\XXint\displaystyle\textstyle{#1}}%
   {\XXint\textstyle\scriptstyle{#1}}%
   {\XXint\scriptstyle\scriptscriptstyle{#1}}%
   {\XXint\scriptscriptstyle\scriptscriptstyle{#1}}%
   \!\int}
\def\XXint#1#2#3{{\setbox0=\hbox{$#1{#2#3}{\int}$}
     \vcenter{\hbox{$#2#3$}}\kern-.5\wd0}}

\def\dashint{\Xint-}

\usepackage{stmaryrd}
\usepackage{multirow}
\usepackage{bigints}

\begin{document}

\title{Electrohydrodynamic drift of a drop away from an insulating wall}
%migration of a droplet near a boundary}
%lift of a drop from an insulating wall}
%migration of a drop away from an insulating wall}
%/Boundary effects on drop electrohydrodynamics}
% \author{ Diptendu Sen$^1$, Mohammadhossein~Firouznia$^{2,3}$, Jeremy A.~Koch$^4$, David Saintillan$^2$, Petia M.~Vlahovska$^4$}
% \affiliation{%
% $^1$ Department of Mechanical Engineering, Northwestern University, Evanston, IL, 60208, USA\\
% $^2$ \textit{Department of Mechanical and Aerospace Engineering,  University of California San Diego, \\
% 9500 Gilman Drive, La Jolla, California, 92093, USA}\\
% \vspace{0.0cm}
% $^3$ \textit{Center for Computational Biology, Flatiron Institute, New York, New York, 10010, USA}\\
% $^4$ Department of Engineering Sciences and Applied Mathematics, Northwestern University, Evanston, IL, 60208, USA
% }
\author{Diptendu Sen}
\thanks{These two authors contributed equally.}
\affiliation{Department of Mechanical Engineering, Northwestern University, Evanston, IL, 60208, USA}
\author{Mohammadhossein~Firouznia}
\thanks{These two authors contributed equally.}
\affiliation{Department of Mechanical and Aerospace Engineering,  University of California San Diego, \\
9500 Gilman Drive, La Jolla, CA, 92093, USA}
\affiliation{Center for Computational Biology, Flatiron Institute, New York, NY, 10010, USA}
\author{Jeremy A.~Koch}
\affiliation{Department of Engineering Sciences and Applied Mathematics, Northwestern University, Evanston, IL, 60208, USA}
\author{David Saintillan}
\affiliation{Department of Mechanical and Aerospace Engineering,  University of California San Diego, \\
9500 Gilman Drive, La Jolla, CA, 92093, USA}
\author{Petia M.~Vlahovska}
\affiliation{Department of Engineering Sciences and Applied Mathematics, Northwestern University, Evanston, IL, 60208, USA}

\date{\today}
%\pacs{47.52.+j, 47.15.G-, 47.55.D-, 47.55.N-, 47.65.Gx}

\begin{abstract}
%A uniform electric field induces an axisymmetric flow about a drop suspended in an unbounded fluid. 
An isolated charge-neutral drop suspended in an unbounded medium does not migrate in a uniform DC electric field.
 A nearby wall breaks the symmetry and causes the drop to drift towards or away from the boundary, depending on the electric properties of the fluids and the wall. In the case of an electrically insulating wall and an electric field applied tangentially to the wall, the interaction of the drop with its electrostatic image gives rise to repulsion by the wall. However, the electrohydrodynamic flow causes either repulsion for a drop with $\mathrm{R/P}<1$, where $\mathrm{R}$ and $\mathrm{P}$ are the drop-to-medium ratios of conductivity and permittivity, respectively, or attraction for $\mathrm{R/P}>1$. 
We experimentally measure droplet trajectories and quantify the wall-induced electrohydrodynamic lift in the case $\mathrm{R/P}<1$. Numerical simulations using the boundary integral method agree well with the experiment and also explore the $\mathrm{R/P}>1$ case. The results show that the lateral migration of a drop in a uniform electric field applied parallel to an insulating wall is dominated by the long-range flow due to the image stresslet.

%{\color{blue}{(Note: I think we can mention 'parallel to the wall' here as well)}}

%The electrohydrodynamic is longer ranger drop velocity decreases more slowly compared to the velocity due to the dielectrophoretic interaction
%and demonstrate that it is well-described theoretically by an image stresslet related to the drop polarization. This migration speed scales inversely with the square of the distance to the wall, a stronger dependence than dielectrophoretic migration. However, for drops with a large drop-to-medium viscosity ratio $\Lr>1$, the electrohydrodynamic flow is weakened and dielectrophoretic migration becomes dominant.
\end{abstract}

\maketitle

\section{Introduction}

Electric fields are widely used to manipulate particles such as colloids \cite{Velev2009,BOYMELGREEN2022,Diwakar2022,Harraq2022},  droplets \cite{GANANCALVO2018,Vlahovska2019}, cells and cellular mimetics \cite{McCaig2009,Messerli2011,Agrawal2022} 
or to assess properties of biomimetic membranes \cite{Dimova-Aranda2007,Aleksanyan2023}. In many of these applications, the particles are close to boundaries. In an applied uniform electric field, even a charge-free particle can drift due to the interaction of the particle induced dipole and its image \cite{JonesTB,Yariv2006,Kilic2011, Wang2022,Wang2023}. This dielectrophoretic (DEP) interaction is attractive with an electrode, and repulsive with an insulating wall. In the far field,  the induced migration velocity decreases with the inverse fourth power of the distance to the boundary. 

Electrohydrodynamic (EHD) flows induce 
%a stronger 
interaction that is  longer-ranged compared to the DEP one, decaying in the far field  as the inverse second-power of the distance to the boundary. To understand this interaction, it is useful to consider an analogy with the 
%quadrupolar
induced-charge electroosmotic (ICEO) flow around a polarizable particle, where fluid  is drawn along the field axis and expelled radially in the equatorial plane. Near 
%an insulating 
a wall, this flow pumps fluid into the gap between the particle and the wall, creating an effective repulsion \cite{Zhao_Bau2007,Kilic2011}. 
%involves fluid drawn along the field axis and expelled radially in the equatorial plane 
%is predicted to cause a metal (ideally polarizable) cylinder or a sphere to be repelled from an insulating wall in a DC field \cite{Zhao_Bau:2007,Kilic:2011}. 
%To understand this interaction, it is useful to consider an analogy with ICEO, where
The flow around a droplet in a uniform electric field exhibits a pattern similar to the ICEO \cite{Taylor1966}, and thus a drop is expected to migrate relative to a nearby wall. However, unlike an ideally polarizable particle, the direction of droplet migration depends on the electric properties of the droplet and suspending fluid. The electrohydrodynamic flow is driven by electric shear stresses due to induced surface charges \cite{Taylor1966, Melcher-Taylor1969}. For a drop in an unbounded medium and subjected to  a uniform electric field, the resulting flow is axisymmetric
about the applied field direction. In the case of a spherical
drop,
%with radius $a$  placed in an \colb{applied uniform} DC electric field $\bm E_{0}=E_0\zhat$,
the interfacial velocity is
\begin{equation}
\label{ehdUsurf}
\bm u\left(r=a,\theta,\phi\right)=\beta_T \sin 2\theta\,\that,
\end{equation}
where ($r,\theta,\phi$) are the spherical coordinates, with $\theta$ being the angle away from the applied field direction, $\bm E_{0}=E_0\zhat$, and 
\begin{equation}
\beta_T=\frac{9 \eps_\out E^2_0 a  }{10 \mu_\out}\frac{\mathrm{R-P}}{\left(1+\mathrm{\lambda}\right)\left(\mathrm{R}+2\right)^2}\,\,,
\end{equation}
 $a$ is the drop radius, and 
$\visrat=\mu_\ins/\mu_\out$ is the viscosity ratio between the drop  and suspending fluid.
%and $\theta$ is the angle with the applied field direction.
%The direction of the surface flow depends on the  difference of conductivity, $\sigma$, and permittivity, $\eps$, of the drop and suspending fluids $\mathrm{R}=\sigm_\ins/\sigm_\out$ and $\mathrm{P}=\eps_\ins/\eps_\out$.  
The direction of the surface flow is determined by the difference of the conductivity ratio ($\mathrm{R}=\sigm_\ins/\sigm_\out$) and the permittivity ratio ($\mathrm{P}=\eps_\ins/\eps_\out$) between the drop and the suspending fluid.
If $\mathrm{R/P}<1$, the  surface  flow is from pole to equator, i.e., the fluid is drawn in at the poles and pushed away from the drop at the equator. The flow direction is reversed for  $\mathrm{R/P}>1$. Accordingly, by analogy with the ICEO effect on a particle near an insulating wall,  the EHD flow is expected to repel a droplet with $\mathrm{R/P}<1$ and attract one with $\mathrm{R/P}>1$.

%The droplet migration velocity can be estimated by considering the 

To estimate the droplet migration velocity, we consider the far field of the unbounded EHD flow, which is a stresslet flow,
%The stresslet strength is 
\begin{equation}
\label{ehdUs}
\bm u\left(\bm r\right)=\beta_T \left(-1+3\cos^2\theta\right)\frac{a^2}{r^2}\bm{\hat{r}}\,.
\end{equation}
To satisfy the boundary conditions at the wall, a reflection to this velocity is introduced.
%such that the sum satisfies the appropriate boundary conditions
%In a flow bounded by a plane, the disturbance velocity produced by a droplet will not satisfy boundary con- ditions on the surface of the plane. As a consequence, we must introduce a reflection to this disturbance velocity such that the sum satisfies the appropriate boundary conditions.
Thus,  for a droplet far from the plane, the leading order droplet migration  velocity
%induced by the presence of the plane 
is  that of the flow velocity of the stresslet image system  evaluated at the position of the droplet;
%Far-away from the drop, in an unbounded fluid the flow 
%is axisymmetrically aligned with the applied field direction and it 
%is approximated by the flow due to an axisymmetric stresslet aligned with the applied field direction. 
%The stresslet strength is 
%A force--free and torque--free particle far from the wall 
%The droplet moves with a velocity due to its corresponding image stresslet; 
in particular, the drop migration velocity $U_{EHD}$ normal to a rigid wall due to the electrohydrodynamic flow is proportional to the stresslet component $S_{nn}$ in the direction of the plane unit normal \cite{Smart-Leighton1990},
\begin{equation}
  U_{EHD}=-\frac{9 S_{nn}}{64\pi\mu_\out h^2},
\end{equation}
where $h$ is the distance from the droplet center to the wall. In the configuration described in Fig.~\ref{fig:hovering_drop},  $S_{nn}= 8 \pi \mu_\out \beta_T a^2 / 3$ and the corresponding migration velocity is
%the applied electric field, $E_0$, is perpendicular to the wall normal, and the corresponding migration velocity of a drop with radius $a$ can be found out as
\begin{equation}\label{u_EHD_dim}
    U_{EHD}=-\frac{1}{h^2} \left(\frac{\eps_\out E^2_0 a^3}{\mu_\out}\right)
    \left[\frac{27(\mathrm{R-P})}{80(1+\mathrm{\visrat})(2+\mathrm{R})^2} \right].
\end{equation}
The velocity induced by the electrohydrodynamic flow  decays more slowly compared to the migration velocity due to the DEP force \cite{Vlahovska2019,Chiara2020}, which is given by
\begin{equation}\label{u_DEP_dim}
  U_{DEP}=\frac{1}{h^4} \left(\frac{\eps_\out E_0^2 a^5}{\mu_\out}\right)\bigg[\frac{3(1+\Lr)}{8(2+3\visrat)}\left(\frac{1-\mathrm{R}}{2+\mathrm{R}}\right)^2\bigg].
\end{equation}
Note that Eqs.~(\ref{u_EHD_dim}) and (\ref{u_DEP_dim}) are far-field descriptions of the drop velocity in the limit of no deformation and negligible surface-charge convection. 

In this paper, we test these theoretical predictions using a combination of experiment and numerical simulations.
%In the subsequent sections, we systematically examine the wall-induced migration of a silicone oil droplet in a castor oil medium, under the influence of a uniform DC electric field, through both experimental investigations and numerical simulations.

\section{Materials and methods}
%{\col{GENERAL REMARK: use present tense! }}
\subsection{Experiments}
\begin{figure}[t]
\centering
\includegraphics[width=0.4\textwidth]{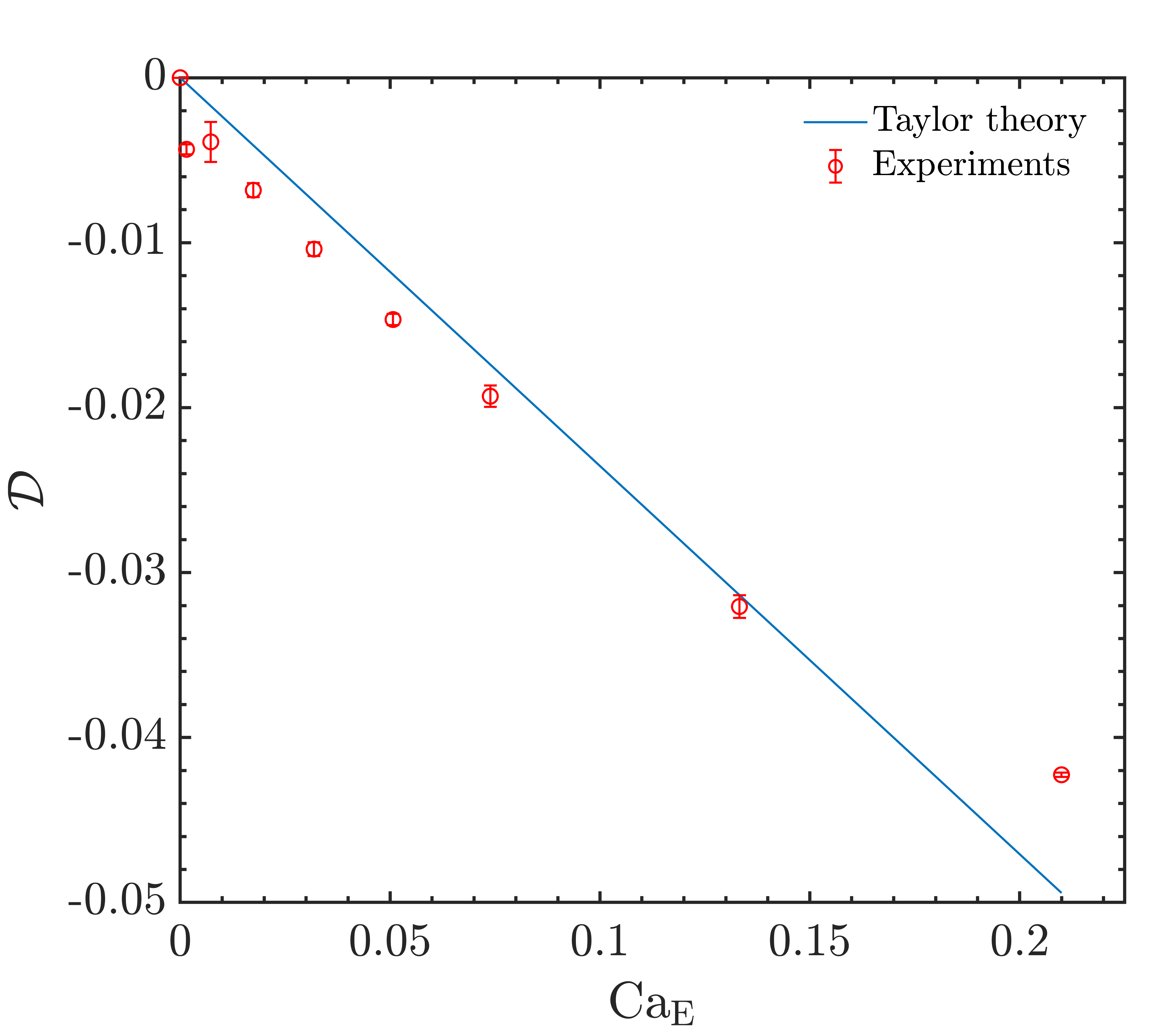}
\caption{\textbf{Oblate drop deformation as a function of the electric capillary number}. The red markers show the experimental results, and the blue line is from Taylor's small deformation theory \cite{Taylor1966}. The error bar shows the standard deviation from three experiments. The droplet radius is $a=1.8$~mm.}
%{\col{what is the error bar? is this results from only one drop? i thought you would put many experiments here and the error bars will be the averages from the experiments. 1 experiment is not very convincing :) }}}
\label{fig:taylor_theory}
\end{figure}

\begin{table*}[t]
  \centering
  \begin{tabular}{lcccccc}
      \hline
      Material & Density, $\rho$& Relative permittivity, $\eps_r$ & Conductivity, $\sigma$ & Viscosity, $\mu$ & \ $t_{EHD}$, $\frac{\mu}{\eps E_0^2}$ & $t_{c}$, $\frac{\eps}{\sigma}$\\
      & (kg/m$^3$) & (-/-) & (pS/m) & (Pa s) & (s) & (s)\\
      \hline
      Silicone oil & 961.08  & 1.70  &  1.2  &  0.047 & - & 12.54\\
      Castor oil   & 958.46  & 2.0  &  45   &  0.66 & 0.25-2.0 & 0.39\\

      \hline
      & $\Delta\rho=2.62~\text{kg/m}^3$& $\mathrm{P}=0.85$ & $\mathrm{R}=0.027$ & $\mathrm{\lambda}=0.07$ & - & -\\
    \hline
  \end{tabular}
  \caption{Material properties for the leaky dielectric drop (silicone oil) and medium (castor oil) used in this study.
  %-- most relevant to this study are the drop-to-medium property ratios: $\Sr=\eps_\text{d}/\eps$, $\Rr=\sigma_\text{d}/\sigma$, and $\Lr=\mu_\text{d}/\mu$.
}
  \label{tab:materials}
\end{table*}
%The density of the silicone oil is reported by Sigma-Aldrich as 0.96~g/mL -- a more accurate density is approximated here based on the sedimentation speed of
A polydimethylsiloxane (PDMS, hereafter referred to as silicone oil) drop is immersed in a castor oil medium, both obtained from Sigma-Aldrich. The material properties are summarized in Table~\ref{tab:materials}. The density of the fluids is measured directly from the weight of a known volume of the liquids. Viscosity values are measured with a TA Instruments Discovery HR-30 rheometer. Permittivity values are measured using a rheo-impedance spectroscopy stage and a 40 mm parallel plate for a Discovery HR-30 rheometer obtained from TA Instruments connected to a Keysight E4990A Impedance Analyzer. Conductivity values were measured in a previous work \cite{sal2010ehd} and the reported values correspond
%Notably this system has
to permittivity ratio $\mathrm{P}=0.85$, conductivity ratio $\mathrm{R}=0.027$ and viscosity ratio $\mathrm{\lambda}=0.07$. 
%It will be shown that the drop migration for this system is dominated by electrohydrodynamic forcing ($U\sim 1/h^2$). 
The material properties are validated by comparing the steady-state shape deformation of a silicone oil drop in castor oil with Taylor's small deformation theory \cite{Taylor1966} in the limit of weak electric fields, see Fig.~\ref{fig:taylor_theory}. {The deformation parameter $\mathcal{D}$ is used to quantify the drop's deviation from a spherical shape and is defined as:
\begin{equation}
\mathcal{D}=\dfrac{l-b}{l+b},
\end{equation}
where $l$ and $b$ denote the drop's major axes parallel and perpendicular to the applied electric field, respectively.}
%As shown in Figure~\ref{fig:taylor_theory}, the experiments align closely with the theory, confirming the accuracy of our measurements.

The droplet dynamics in the electric field involves processes occurring on different time scales.
%We introduce next the various timescales relevant to the problem at hand. 
Of particular interest is the timescale of the electrohydrodynamic flow $t_{{EHD}}={\mu_\out}/{\eps_\out E_0^2}$. The conduction of charges is dictated by the charge relaxation time $t_c={\eps_\out}/{\sigma_\out}$, while the visco-capillary timescale $t_{\gamma}={\mu a}/{\gamma}$ governs the drop shape relaxation. {The comparison of  these time scales defines the dimensionless parameters
\begin{equation}
	\mathrm{Ca_\mathrm{_E}}=\frac{t_{\gamma}}{t_{{EHD}}}=\frac{\varepsilon \, a \, {E^2_{0}} }{\gamma}, \qquad \mathrm{Re_\mathrm{_E}} =\frac{t_{c}}{t_{{EHD}}}=\frac{\varepsilon^2 \, {E^2_0}}{ \mu \, \sigma}.  \label{eq:nondim:Ca}
\end{equation} 
The electric capillary number $\mathrm{Ca_\mathrm{_E}}$ measures the relative strength of electric forces against surface tension, while the electric Reynolds number $\mathrm{Re_\mathrm{_E}}$ determines the significance of surface charge convection relative to Ohmic conduction.} 

The silicone oil/castor oil system is commonly used in electrohydrodynamic studies due to the relatively small density difference between the two materials, {with ${\Delta\rho}/{\rho}\sim O(10^{-3}).$} 
%{\col{better show $\Delta\rho/\rho$ to show that indeed the desnity deifference is small } }. 
This density difference results in drop migration due to buoyancy on a time scale that is significantly longer than the electrohydrodynamic time scale in our system, with $t_{{EHD}}/t_{b} \sim O(10^{-2})$ at most.  The buoyancy time scale, $t_b=\frac{3 \mu}{2ag \Delta \rho}\left(\frac{2+3\mathrm{\lambda}}{1+\mathrm{\lambda}}\right)$, is estimated from the Hadamard-Rybczynski settling velocity \cite{batchelor2000introduction}. Finally, the inertial-viscous Reynolds number, $\mathrm{Re}=\rho a^2 \varepsilon E^2_{0}/\mu^2$,  is estimated, using typical material properties and experimental electric field strengths, to be $\mathrm{Re}\sim O(10^{-3})$. Consequently, inertial effects are negligible in this system.

%We estimate the gravitational forcing to be negligible compared to the EHD force, which will become evident from our results in figure~\ref{fig:nondim}. 

%The properties were validated against the steady state shape deformation (silicone oil drop in castor oil) and Taylor's small deformation theory~\cite{Taylor:1966} for weak electric fields. The results are shown in figure~\ref{fig:taylor_theory}. The reasonable agreement of the experiments with Taylor's theory validates our experimental setup.

%\subsection{Experimental setup}
The experimental setup consists of a rectangular chamber formed by four glass plates mounted in a 3D-printed base. Two opposing sides ($75~\text{mm}\times50~\text{mm}$) 
are made 
%conductive on one side with an 
from indium-tin oxide (ITO) glass (Delta Technologies), allowing the plates to serve as transparent electrodes. The other two plates ($75~\text{mm}\times25~\text{mm}$) are nonconductive, providing visual access to the experiment and create $d=25$ mm spacing between the electrodes. A rectangular Teflon insert (thickness $12.7~\text{mm}$) is placed at the base of the chamber, serving as the insulating wall. Attached to the ITO-coated glass are high-voltage wires connected to an Ultravolt 40A12-P4 high-voltage converter powered by an Agilent E646A DC power supply.

\begin{figure}[t]
\centering
\includegraphics[width=0.85\textwidth]{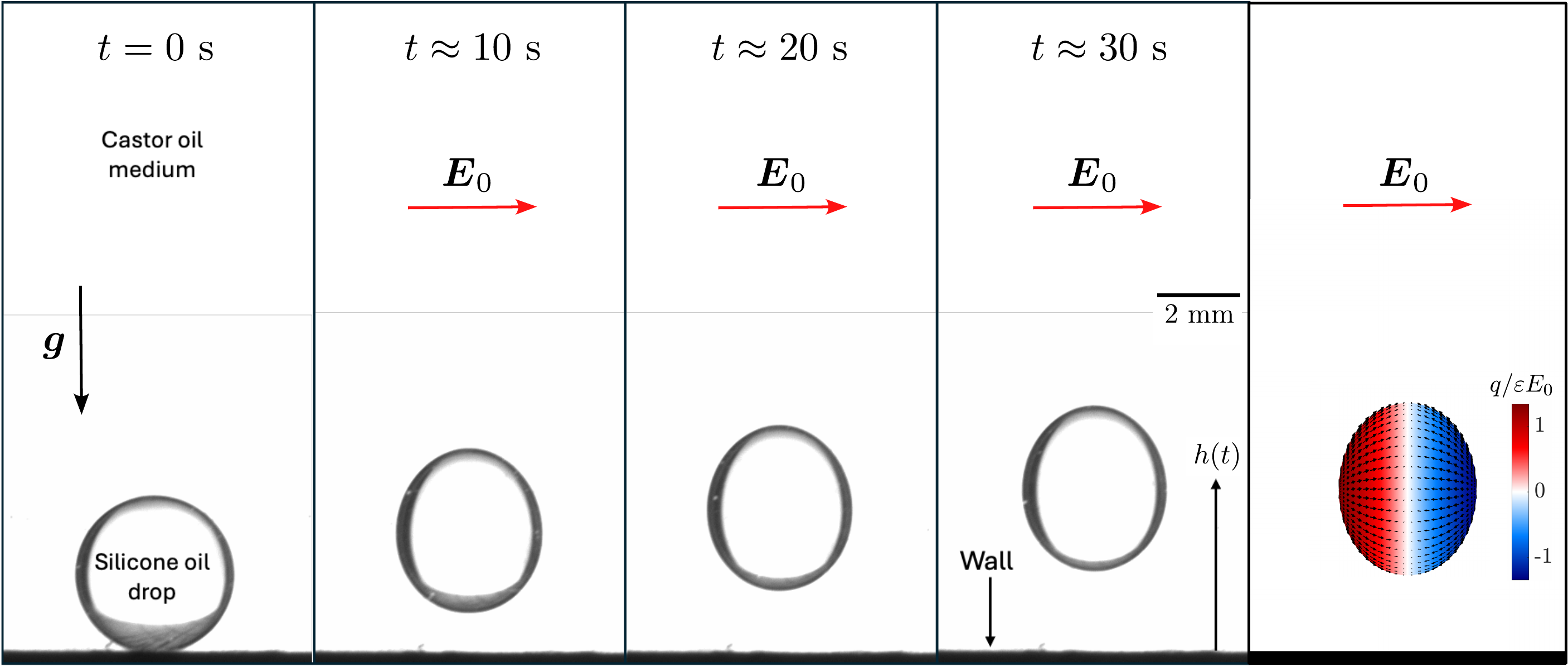}
\caption{\textbf{Snapshots of a droplet migrating away from a horizontal boundary}. The figure shows experimental images (panels 1-4) and a simulation result (last panel on the right) for a drop with $(\mathrm{R},\,\mathrm{P},\,\mathrm{\lambda})=(0.027,\, 0.85,\, 0.07)$ migrating away from an insulating wall in an  electric field, applied tangentially to the wall, corresponding to an electric capillary number $\mathrm{Ca_{_E}}=0.43$. The color on the drop from the numerical simulations represents the induced surface charge, while the overlaid vectors illustrate the interfacial velocity field.}
\label{fig:hovering_drop}
\end{figure}

In the experiments, a constant DC voltage $V$ is applied 
%across the container's thickness $d=25~\text{mm}$ 
to generate a uniform electric field of strength $E_0=V/d$. In this study, the strength of electric field is varied between $1.37 \, \mathrm{kV/cm}$ and $3.8 \, \mathrm{kV/cm}$.
The trajectories of the drop are recorded using a Thorlab DCU224M camera with Navitar Zoom 7000 lens collecting images at 15~frames per second.  %{\st{In these images, the view direction is perpendicular to both the applied electric field and the wall normal. From the camera's perspective, the electric field points left to right, gravity acts downward, and the wall normal points upward.}} : {\col{This sentence may be unnecessary, just put the gravity g direction on the plot}}

The experiment is prepared by injecting a silicone oil drop of radius $a\sim1.5\text{~mm}-2\text{~mm}$ into the chamber filled with castor oil. Due to the slight density mismatch in the materials (Table \ref{tab:materials}), the drop slowly sediments towards the Teflon base. When it gets close to the base, at a distance $h_0$, the signal generator is switched on to apply the voltage, providing a  constant DC electric field. The ensuing motion of the drop is monitored for approximately $30$~seconds and saved as a video.

Fig.~\ref{fig:hovering_drop} shows snapshots of a drop drifting away from the wall. The application of the electric field causes the drop to deform oblately, taking the appearance of an ellipse from the perspective of the camera. The droplet trajectories are calculated using \textsc{MATLAB} to track the position of the ellipse center as a function of time, creating a data set of the drop height relative to the wall: $h$~(mm), as a function of time, $t$~(s). Time $t=0$ is defined as the frame in which the drop just begins to lift off from the wall.

%Time $t=0$ is defined as the frame in which  the \textit{E}-field is applied {\color{OliveGreen} Our t = 0 for the plots is when the drop just begins to lift off from the wall}.

\subsection{Numerical simulations}

{This section presents the governing equations used in the numerical simulations, expressed in nondimensional form. The characteristic quantities used for nondimensionalization are length $a$, time $t_{c}=\eps/\sigma$, velocity $a/t_{{EHD}}$, pressure $\eps E_0^2$, charge $\varepsilon E_0$, and electric potential $a\,E_0$. All equations presented hereafter are expressed in these dimensionless terms.} We consider a neutrally buoyant drop of leaky dielectric fluid occupying a volume $V^-$, immersed in a semi-infinite body of another leaky dielectric fluid $V^+$, positioned near a flat wall. The system is subject to a uniform electric field oriented parallel to the wall. Following the Taylor--Melcher leaky dielectric model \citep{Melcher-Taylor1969}, we assume that any free charge in the system is confined to the interface $\partial V$, and the bulk of the fluids remains electroneutral. Consequently, the electric potential within the bulk is governed by Laplace's equation. The electric problem can be formulated in an integral form \citep{sherwood1988breakup, baygents1998EHD, lac2007axisymmetric}. For every $\bm{x}_0\in V^{\pm},\, \partial V$:
\begin{equation}\label{eq:BEM:potential_SLP}
    \begin{aligned}
    \varphi( \bm{x}_0 )=-\bm{x}_0\bcdot\bm{E}_{0} -\int_{\partial V} \bm{n}\bcdot\llbracket \nabla \varphi( \bm{x} ) \rrbracket \,\mathcal{G}^w \left( \bm{x}_0;\bm{x} \right) \, \mathrm{d}s(\bm{x}),
    \end{aligned}
\end{equation}
where $\mathcal{G}^w \left( \bm{x}_0;\bm{x} \right)=(4\pi r)^{-1}+(4\pi \tilde{r})^{-1}$ is the Green's function describing the electric potential due to a point charge near an insulating wall located at $x_w=0$. Here, $\bm{\tilde{x}}_0$ is the mirror image of $\bm{x}_0$ with respect to the wall, with  $\bm{r} = \bm{x}_0-\bm{x}, ~r=|\bm{r}|$ and $\bm{\tilde{r}} = \bm{\tilde{x}}_0-\bm{x}, ~\tilde{r}=|\bm{\tilde{r}}|$. The operator $\llbracket g \rrbracket \coloneqq g^+- g^-$ denotes the jump in any variable $g$ across the interface $\partial V$. According to Gauss's law, the surface charge density is related to the jump in the normal electric field across the interface as $q(\bm{x})=\bm{n} \bcdot  (\bm{E}^+ - \mathrm{P}\bm{E}^-)$, where $\bm{x}\in \partial V$. Taking the gradient of Eq.~\eqref{eq:BEM:potential_SLP} with respect to $\bm{x}_0$ and considering the jump across the interface, we derive an integral equation for the jump in the normal electric field. For $\bm{x}_0\in \partial V$:
  \begin{equation}
\frac{1}{2}\llbracket E^{n}(\bm{x}) \rrbracket = E^{n}_{0}(\bm{x}_0) - \dashint_{\partial V} \! \! \llbracket E^{n}(\bm{x}) \rrbracket [\bm{n}(\bm{x}_0)\bcdot\bnabla_0 \mathcal{G}^w]\,\mathrm{d}s(\bm{x}).\label{eq:BEM:integral_eq_En}
\end{equation}
%{\col{$Re_E$ is not defeined. Specify somwhere that all variables are nondimensionalize withe time scale ?, drop radius and field magnitude.  }}
The surface charge evolves due to bulk Ohmic and convective surface currents, satisfying the conservation equation:
\begin{equation} 
	 \partial_t q+\bm{n}\bcdot\left( \bm{E}^+ - \mathrm{R} \bm{E}^- \right) + \mathrm{Re_{_E}}\bnabla_s \bcdot(q \bm{u})=0, \qquad \bm{x}\in \partial V. \label{eq:pb_def:chgcons}
\end{equation}
where $\bnabla_s = (\bm{I}-\bm{nn})\bcdot \bnabla$ denotes the surface gradient operator.

In the absence of inertial and buoyancy effects, the velocity and pressure fields are governed by the Stokes and continuity equations. The flow problem is then recast as a boundary integral equation \cite{rallison1978numerical, pozrikidis1992BIM_book}. For every $\bm{x}_0\in \partial V$:
\begin{equation}
    \begin{aligned}
    \bm{u}(\bm{x}_0) =  - \mathcal{K} \int_{\partial V}    \llbracket \bm{f}^\mathrm{H}(\bm{x}) \rrbracket  \bcdot \bm{G}^w(\bm{x}_0;\bm{x})\,\mathrm{d}s(\bm{x}) 
		 + (\mathrm{ {1-\lambda}})\mathcal{K} \,\, \dashint_{\partial V}\bm{u}(\bm{x})\bcdot\bm{T}^w(\bm{x}_0;\bm{x}) \bcdot \bm{n}(\bm{x})\, \mathrm{d}s(\bm{x}),
    \end{aligned}
\end{equation}
where $\mathcal{K}=1/(4\pi\mathrm{(1+\lambda)} )$. Here, $\bm{G}^w(\bm{x}_0;\bm{x})$ is Blake's Green's function for the flow due to a unit point force near a plane wall, and $\bm{T}^w(\bm{x}_0;\bm{x})$ is the corresponding stress tensor \cite{blake1971note,blake_chwang1974viscous_flow_singularities}. 

The balance of external and internal forces at the interface is governed by:
\begin{equation}
	\llbracket \bm{f}^\mathrm{H}\rrbracket + 
    \llbracket \bm{f}^\mathrm{E}\rrbracket=\mathrm{Ca^{-1}_{_E}}\, (\bnabla_s\bcdot \bm{n})\,\bm{n},
    \qquad \bm{x}\in \partial V. \label{eq:pb_def:dyn_BC}
\end{equation}
This dynamic boundary condition ensures that the jump in hydrodynamic and electric tractions is balanced by capillary forces, assuming uniform surface tension ($\bnabla _s \gamma = \bm{0}$). Hydrodynamic and electric tractions are expressed in terms of the Newtonian and Maxwell stress tensors, respectively:
\begin{align}
\!\!\! \llbracket \bm{f}^\mathrm{H} \rrbracket &= \bm{n} \bcdot \big[ 
    \big(-p \bm{I} + (\bnabla\bm{u} +{\bnabla\bm{u}}^T)\big)^+ 
    - \big(-p \bm{I} + \lambda (\bnabla\bm{u} +{\bnabla\bm{u}}^T) \big)^- 
\big], \label{eq:pb_def:jump_fH} \\
\!\!\!  \llbracket \bm{f}^\mathrm{E} \rrbracket &= \bm{n} \bcdot \big[ 
    \big(\bm{E}\bm{E} - \tfrac{1}{2}E^2\bm{I}\big)^+ 
    - \mathrm{P} \big(\bm{E}\bm{E} - \tfrac{1}{2}E^2\bm{I}\big)^- 
\big], \label{eq:pb_def:jump_fE}
\end{align}
Finally, the drop's shape evolves according to the normal velocity at the interface:
\begin{equation}
 \partial_t{\bm{x}}=\mathrm{Re_{_E}}(\bm{u}\bcdot\bm{n})\bm{n},\qquad \bm{x}\in \partial V. \label{eq:pb_def:kinematic_BC}
\end{equation}
The velocity of the drop's center of volume, $\boldsymbol{U}_{c}$, is calculated as:
\begin{equation}
    \boldsymbol{U}_{c}(t)=\frac{1}{V_\ins}\int_{V^-}\boldsymbol{u} \, \mathrm{d}v= \frac{1}{V_\ins}\int_{\partial V} \boldsymbol{x}\,(\boldsymbol{n\cdot u}) \, \mathrm{d}s, \label{eq:UC_integral}
\end{equation}
where $V_\ins$ is the drop volume. In our system, the velocity is aligned with the $x-$axis, such that $\boldsymbol{U}_{c}= U_c \,\boldsymbol{e}_x$.

\begin{figure}
\centering
\includegraphics[width=0.5\textwidth]{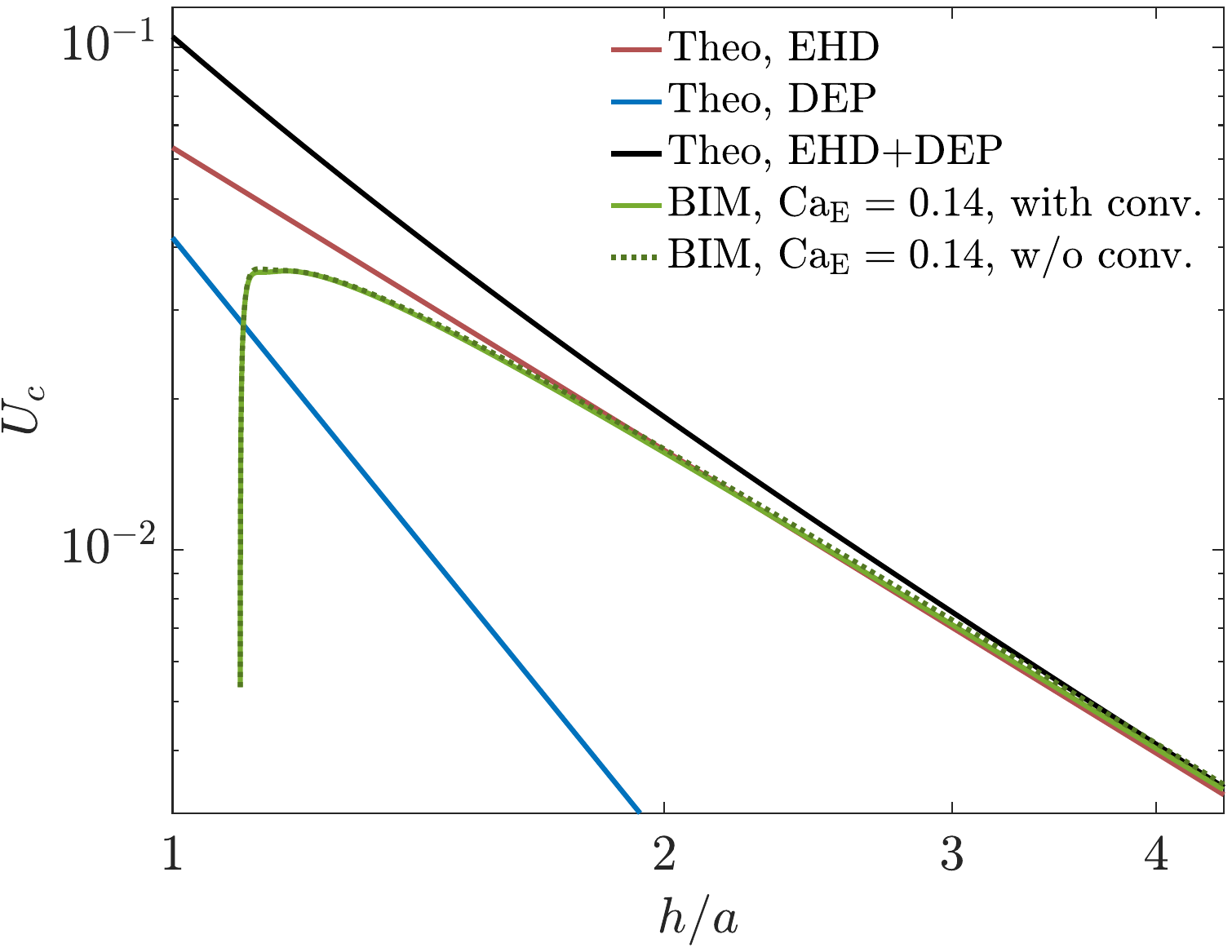}
\caption{ {\bf Comparison between the migration velocity computed numerically and theoretically from Eqs.~\eqref{u_EHD_dim} and \eqref{u_DEP_dim}.} 
%{\bf{Effect of charge convection on droplet migration.}} 
Migration velocity as a function of distance from the wall for a drop with $(\mathrm{R},\,\mathrm{P},\,\mathrm{\lambda})=(0.027,\, 0.85,\, 0.07)$ and  $(\mathrm{Ca_E}, \, \mathrm{Re_E})=(0.14,\, 0.19)$ with and without charge convection. Results from boundary integral simulations are compared with the theoretical predictions from Eqs.~\eqref{u_EHD_dim} and \eqref{u_DEP_dim}. The drop is initially positioned at $h_0/a=1.1$ in the simulations. } 
\label{fig:charge_convection}
\end{figure}

We numerically solve Eqs.~\eqref{eq:BEM:integral_eq_En}--\eqref{eq:pb_def:dyn_BC} and \eqref{eq:pb_def:kinematic_BC} using a spectral boundary integral solver developed for electrohydrodynamic flows in viscous drops \cite{firouznia2023JCP, Firouznia_Spectral_Boundary_Integral_2022}. Fig.~\ref{fig:hovering_drop} compares snapshots of a drop migrating away from the wall in experiments and simulations, with good agreement between the two.
Fig.~\ref{fig:charge_convection} provides a more precise validation of the numerical method. It illustrates the evolution of the droplet migration velocity upon application of the field obtained from simulations. After an initial transient, due to droplet deformation and polarization, the velocity approaches the theoretical result,  \eqref{u_EHD_dim}. 
%that 
We performed two sets of simulations: one that fully accounts for charge conservation as described in Eq.~\eqref{eq:pb_def:chgcons}, and another in which the convective term is neglected.
%by setting $\mathrm{Re_E} = 0$. 
The results show that charge convection has little impact on the  migration velocity.
%at this low $\mathrm{Ca_E}$. 
%, where the asymptotic predictions in \eqref{u_EHD_dim} and \eqref{u_DEP_dim} remain most accurate.
Consequently, charge convection is omitted in the simulations for the remainder of this study. At higher electric field strengths, however, the role of charge convection becomes more pronounced. For oblate drops with low viscosity ratios, it is known that charge convection can generate steep charge gradients, leading to charge density shocks \cite{firouznia2023JCP,PengSchnitzer2024PRF,Lanauze2015} resulting in equatorial streaming \cite{Brosseau2017b,Wagoner2020,Wagoner2021} or electrorotational instabilities \cite{Paul,Ouriemi2014,Vlahovska2016b}. In the experiments, however,  the range of $\mathrm{Ca_{_E}}$ corresponds to values of $\mathrm{Re_{_E}}$  from 0.2 to 1.53, well bellow the conditions for such instabilities to occur.  
%This phenomenon, while significant, falls beyond the scope of the present work.

\section{Results and Discussion}

\subsection{Oblate drops}

\begin{figure}
\centering
\includegraphics[width=0.9\textwidth]{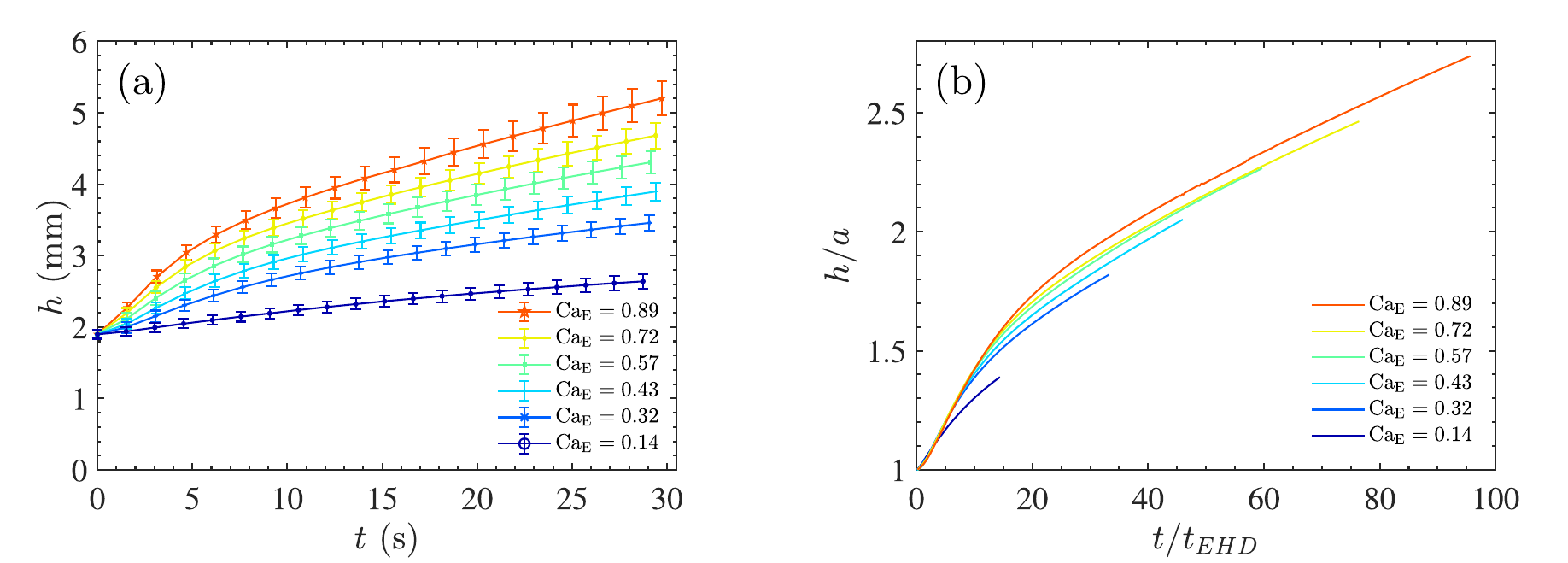}%h_vs_t_avg_P_0p85_diff_marker_final_leg_rev_w_tot_err.png
\caption{{\bf{Droplet drift away from an insulating wall.}}  %The drop migrates away from the wall more quickly at larger electric capillary numbers, $\mathrm{Ca_{_E}} \sim E_0^2$.  
Droplet height above the wall as a function of time at increasing values of the electric field represented by the electric capillary number $\mathrm{Ca_{_E}} =\varepsilon a E_0^2/\gamma$, for a droplet of radius $a=1.90$~mm. The dimensional trajectories are shown in (a), where the lines and error bars show the average migration height and total error from three different experiments combined with the error due to optical distortion and contour detection, respectively. The non-dimensionalized trajectories are shown in (b). 
%The total error in the determination of the radius of the drop is within 6\%.
%{\col{Are these results from several measurements? what is the error bar?}}
}
\label{fig:h_vs_t_dim}
\end{figure}
Fig.~\ref{fig:h_vs_t_dim} shows the drift of the leaky dielectric drop away from the wall, demonstrating the existence of the migration phenomenon and indicating that the  migration speed increases with capillary number (and therefore with electric field strength, $\mathrm{Ca_{_E}}\sim E_0^2$). Fig.~\ref{fig:h_vs_t_dim}(a) shows the dimensional trajectories, while the non-dimensional trajectories are shown in Fig.~\ref{fig:h_vs_t_dim}(b).

In Fig.~\ref{fig:nondim}(a) and (b), the trajectories of two drop sizes $a=$1.46~mm and 1.90~mm, respectively, from experiments, subject to electric fields from $E_0=$1.37~kV/cm to 3.8~kV/cm, are plotted as solid lines on a log-log scale, with the elevation $h$ (distance between the drop center and the wall) and time $t$ nondimensionalized by the drop radius $a$ and by the electrohydrodynamic time $t_{{EHD}}={\mu}/{\epsilon E_0^2}$, respectively. The color indicates the value of the electric capillary number $\mathrm{Ca_{_E}}$, which varies from $\mathrm{Ca_{_E}}=0.1$ to 1. The cube of the initial position is subtracted from the cube of the distance from the wall and has been plotted as a function of time, which 
%ostensibly leaves the scaling independent of 
eliminates the dependence on the initial location. The EHD theory of Eq.~(\ref{u_EHD_dim}) is plotted  as a black dashed line. The long-term behavior of the experimental trajectories is well-matched by the theory; however, the velocity in the near-wall region is overpredicted by Eq.~\ref{u_EHD_dim} -- likely due to near-wall effects not accounted for in the theory, including non-negligible  contributions of higher-order images. The droplet trajectories computed from numerical simulations are also plotted as solid lines in Fig.~\ref{fig:nondim}(c), showing favorable agreement with both the theory and experiments. Experimental trajectories for $\mathrm{Ca_{_E}}=0.32$ and $\mathrm{Ca_{_E}}=0.89$ are also plotted as circular markers in Fig.~\ref{fig:nondim}(c) to demonstrate this agreement. The choice to plot this in a log-log scale was made for two reasons: (1) it more clearly highlights the long term power-law scaling; and (2) it collapses the data for experiments in which the EHD time, which scales with $E_0^2$, varies over a wide range (from 0.25~s to 2~s). In the simulations, the drop is initially positioned at $h/a = 1.1$ with zero charge. Consequently, its early migration is primarily driven by DEP interactions at initial times. As charge accumulates on the drop’s surface, EHD effects intensify, leading to increased migration velocities. During this transient regime, the distance from the wall does not follow the scaling behavior predicted by Eq.~(\ref{u_EHD_dim}) or (\ref{u_DEP_dim}). Eventually, as the charge density approaches its steady distribution and the drop moves farther from the wall, EHD interactions dominate, and the distance from the wall exhibits the
%cubic 
scaling predicted by Eq.~(\ref{u_EHD_dim}).

% -- in linear scaling, the lower \textit{E}-field curves will be significantly shorter.

%The EHD flow as well as the streaming flow past the drop due to its migration could convect the charges induced on the droplet surface. {\color{OliveGreen}The electric Reynolds number is $\mathrm{Re_{_E}}\sim O(10^{-1})$ at lower values of electric capillary numbers $\mathrm{Ca_{_E}}$, the regime where we expect our theory and simulations to be most applicable. Alternatively, the electric Reynolds number can be calculated based on the sedimentation velocity as $\mathrm{Re^*_\mathrm{E}}=\frac{\eps U_{set}}{a \sigma}$, where $U_{set}$ is the Hadamard-Rybczynski settling velocity. We estimate $\mathrm{Re^*_\mathrm{E}} \sim O(10^{-2})$, suggesting that charge convection due to sedimentation is negligible in our system.}

The EHD flow, along with the streaming flow induced by drop migration, sweeps the charges accumulated on the droplet surface. {The extent of the convection effect by the EHD flow is quantified by the electric Reynolds number, which is found to be $\mathrm{Re_{_E}}\approx 0.2$ at the lower values of the electric capillary number, $\mathrm{Ca_{_E}}=0.14$. This regime, characterized by weak charge convection, is where our theory and simulations are expected to be most accurate. 
%The charge convection due to  droplet sedimentation velocity as $\mathrm{Re^*_{_E}}={\eps U_{sed}}/{a \sigma}$, where $U_{sed}$ is the Hadamard-Rybczynski settling velocity. In this case, we estimate $\mathrm{Re^*_{_E}} \sim O(10^{-2})$, indicating that charge convection due to sedimentation is negligible in our system.} 
%{\col{This should be convection due to droplet translation, can you estimate based on the typical value for U from the experiment??}}. 
The charge convection due to the  droplet translation is quantified as $\mathrm{Re^*_{_E}}={\eps U_{t}}/{a \sigma}$, where $U_{t}$ is the droplet translation velocity from our experiments. In this case, we estimate $\mathrm{Re^*_{_E}} \sim O(10^{-2})$, indicating that charge convection due to the streaming flow around the translating drop is negligible in our system.

\begin{figure}
\centering
\includegraphics[width=0.99\textwidth]{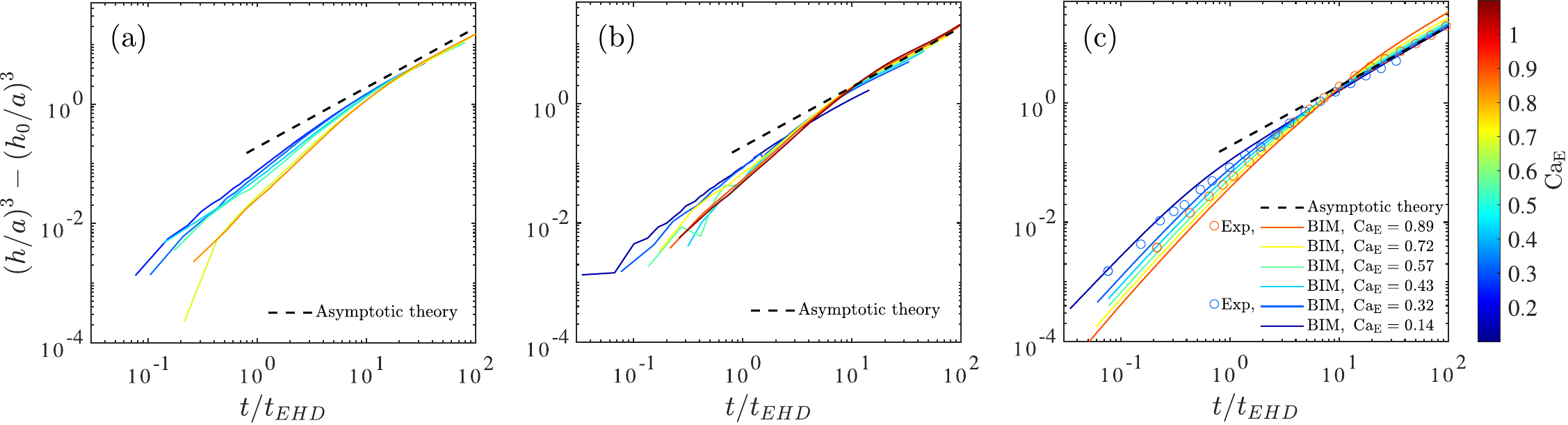}
\caption{ {\bf{Droplet height evolution}}. Experimental measurements of droplet elevation, normalized by the drop radius, plotted as a function of time, normalized by the electrohydrodynamic time,  for drops of radius 1.46~mm and 1.90~mm, are shown in (a) and (b), respectively as solid lines. The color of the trajectories denotes the value of the electric capillary number, which varies from approximately $\mathrm{Ca_{_E}}=0.1$ to 1, showing  favorable agreement with Eq.~(\ref{u_EHD_dim}).  The theory predicts the scaling $(h/a)^3\sim t/t_{{EHD}}$ (black dashed line), emphasizing the dominance of EHD forcing in the far field. The (cube of the) initial height is subtracted from the ordinate to more clearly demonstrate this scaling. The results from boundary integral (BIM) simulations for a drop of radius $1.90$~mm and different capillary numbers are plotted in (c) as solid lines, and agree well with the experiments, plotted as circular markers for $\mathrm{Ca_{_E}}=0.32$ and $\mathrm{Ca_{_E}}=0.89$.} %The inset shows the relative magnitude of contribution of the DEP velocity with respect to the EHD velocity as given by Eq.~(\ref{u_DEP_EHD_ratio}).}
\label{fig:nondim}
\end{figure}

%In our system, however, the charge convection  is negligible. The importance of charge convection is estimated by the ratio of the charge relaxation time to the flow convection time, defining  the electric Reynolds number
 %\begin{equation} \label{reynoldsnumber}
 % Re_\text{E}=\frac{\eps U}{a \sigma}
%\end{equation}
%where U is a characteristic velocity scale. We estimate that $Re_\text{E}$ (based on the maximum EHD velocity at the interface, Eq. \ref{ehdUsurf}) for our system is $\sim O(10^{-1})$, suggesting that charge convection effects can be neglected.
%{\col{CAREFUL: this $Re_\text{E}$ is different that the one in Eq. 9, please clean up the notation. ALSO what is the effect of convection because of the sedimntation velocity? }}
%effects of charge convection on the stresslet strength, following the work of Shutov \cite{shutov2002shape}, and find that the effect on the stresslet velocity for our experimental parameters is, indeed negligible. Gravity effects are also small for our system, especially closer to the wall, and can be neglected.

Fig.~\ref{fig:nondim} demonstrates favorable agreement with Eq.~(\ref{u_EHD_dim}) at long times, which suggests negligible contribution 
%, which predicts an inverse-square relationship between the drop migration velocity and the distance to the wall. 
%Notably, this equation neglects the contribution 
from dielectrophoresis.
%whose associated contribution to the velocity scales as $O(1/h^4)$ with the distance from the wall; see Eq.~\eqref{u_DEP_dim}. Indeed,
The ratio of the two velocity contributions,
\begin{equation}\label{u_DEP_EHD_ratio}
  \frac{U_{DEP}}{U_{EHD}}=
  -\frac{10}{9}\frac{(1+\mathrm{\Lr})^2(1-\mathrm{R})^2}{(2+3\mathrm{\Lr})(\mathrm{R-P})}\bigg(\frac{a}{h}\bigg)^2,
\end{equation}
is, for the present leaky dielectric system, at most 0.66 at $h/a=1$ (i.e., for a drop touching the wall), and is $0.1$ at $h/a=2.5$. 
%The variation of this ratio from Eq.~(\ref{u_DEP_EHD_ratio}) is plotted {\color{blue}{\st{as an inset}}} in Fig.~\ref{fig:U_dep_ehd}, which illustrates this point clearly.
%\begin{figure}
%\centering
%\includegraphics[width=0.40\textwidth]{./Final figures/DEP_to_EHD_P_0p85_main_final.png}
%\caption{Ratio of DEP to EHD velocity as a function of distance from the wall, non-dimensionalised by the drop radius. {\color{red}(suggestion: We can remove this figure entirely.)} }
%\label{fig:U_dep_ehd}
%\end{figure}
However, note that for an electrohydrodynamic system with a smaller value of $(\mathrm{R/P}-1)$ (the factor that sets the EHD flow strength) or a larger drop viscosity (and thereby a larger $\mathrm{\Lr}$; silicone oil, for example, is available at a viscosity 200 times larger than that used here such that $\mathrm{\Lr}\sim14$), the dielectrophoretic force can become consequential in the near-wall migration behavior.

For a mobile interface (i.e., a fluid drop with $\mathrm{\lambda} < \infty$), at distances sufficiently far away from the wall, EHD forcing will eventually overtake DEP forcing, due
to the slower, $1/h^2$, decay of $U_{EHD}$. For a solid ``drop" (particle), $\mathrm{\lambda} \to \infty$ and the EHD flow is supressed. Accordingly, there is no EHD-induced migration, and the migration is purely due to dielectrophoresis. It is worth emphasizing that these equations are far-field descriptions of the flow field -- higher-order descriptions of the flow singularity are unaccounted for, though these contributions are expected to be significant at close to the wall $h/a\sim O(1)$. Despite these approximations, Fig.~\ref{fig:nondim} demonstrates the existence of the electrohydrodynamic migration phenomenon and its domination over the dielectrophoretic migration in leaky dielectric systems.

%Particularly in view of the repulsive nature of the EHD and DEP forcing, the drop will migrate in such a way that Eqs.~(\ref{u_EHD_dim}) and (\ref{u_DEP_dim}) become useful descriptions of the flow behavior.

%As shown in a previous work \cite{wang2022particle}, an attractive DEP force would exist in between a particle(or drop) and the electrode walls in our system. Now, for an oblate system where the EHD and DEP forces (resulting from the bottom insulating wall) are antagonistic to gravity, the repulsive forces are balanced out by gravity as the drop reaches a certain height. At this "hovering height", the main driving force is the DEP interaction with the side walls (or electrodes), and the drop would be attracted to the closer electrode. Indeed, we see in our experiments, as the drop reaches close to the "hovering height", it starts drifting towards the nearest electrode.

%We also explore the drop migration phenomenon for a Prolate drop system numerically, with $R=0.25$, $P=0.063$ and $\lambda=10$. For such a system, the EHD forcing is attractive in nature, while DEP is still repulsive, which suggests there might be a hovering height where the two forces are balanced. The drop migration velocity for such a system is plotted in Figure~\ref{fig:U_Prolate_hovering} as a function of the non-dimensional height, which shows the existence of this hovering height. Although, in experiments, prolate drops tend to break up when $Ca_E$ reaches a value of 0.5. However, such a system could be interesting to explore in future.

\subsection{Prolate drops}

We extend our analysis to prolate drops, characterized by $\mathrm{R/P} > 1$. In this regime, EHD and DEP effects act in opposition, potentially leading to a steady hovering state where the two forces balance. However, prolate systems are harder to realize experimentally, especially since they are unstable and prone to breakup at high electric field strengths needed to induce significant migration. {%\color{OliveGreen}Due to experimental constraints, we do not have direct measurements for this regime. (@Diptendu: Specify the experimental challenges and explain why suitable fluids with the required material properties are unavailable in this regime.)} 
Therefore, our analysis in this section relies on asymptotic theory and numerical simulations. Fig.~\ref{fig:U_Prolate_hovering} presents the magnitude of the migration velocity as a function of wall separation for a prolate drop. Solid lines indicate motion away from the wall, while dotted lines denote motion toward it. As described by Eqs.~\eqref{u_EHD_dim} and \eqref{u_DEP_dim}, DEP and EHD interactions counteract each other: short-range DEP repulsion and long-range EHD attraction establish a steady hovering state at a predicted equilibrium distance of $h/a\approx3.5$ from the wall. We note that the hovering position observed in numerical simulations deviates from theoretical predictions. Additionally, for a viscosity ratio of $\lambda=1$, no hovering state is observed, and the drop instead migrates toward the wall. These discrepancies arise from near-wall hydrodynamic effects, which are not accounted for in the asymptotic theory. Furthermore, increasing the electric capillary number $\mathrm{Ca_{_E}}$ shifts the hovering point further from the wall.

\begin{figure}[t]
\label{u_Prolate}
\centering
\includegraphics[width=0.5\textwidth]{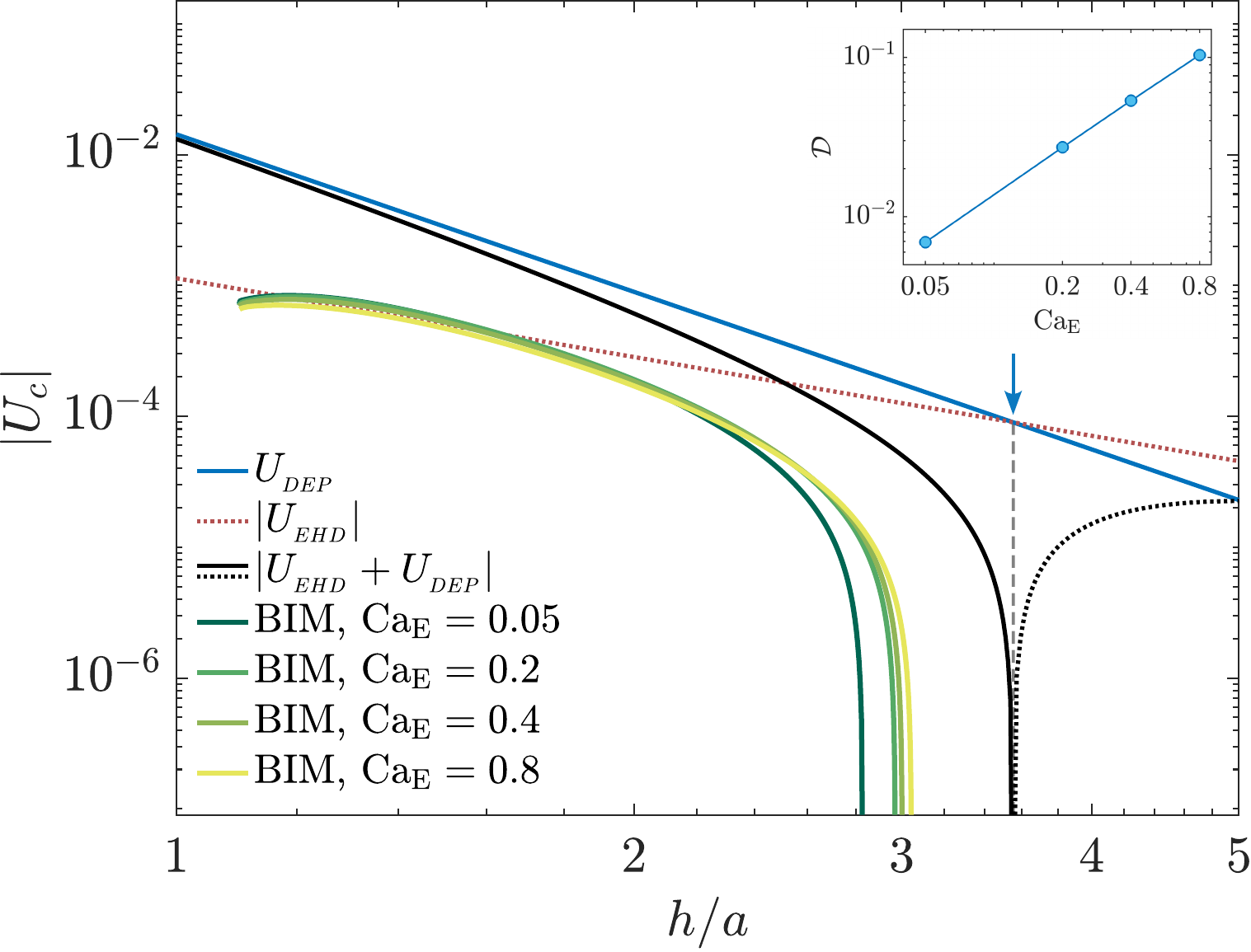}
\caption{{\bf{Antagonistic EHD and DEP interactions can result in a hovering drop}}. Migration velocity as a function of distance from the wall in a drop with $(\mathrm{R},\,\mathrm{P},\,\mathrm{\lambda})=(0.25,\, 0.063,\, 10)$ at various electric capillary numbers, compared with theoretical predictions from Eqs.~\eqref{u_EHD_dim} and \eqref{u_DEP_dim}. In all simulations, the drop is initially positioned at $h_0/a=1.1$ from the wall, and its trajectory is subsequently computed. Solid lines indicate positive migration velocities (away from the wall), while dotted lines indicate negative velocities (towards the wall). The inset shows the deformation parameter at the hovering point in each case. The arrow indicates the hovering height determined by the theory.} 
%{\col{Could you put an arrow or verical dot-dash line at the point where EHD and DEP intersect, and the total velocity becomes zero?}}}
\label{fig:U_Prolate_hovering}
\end{figure}

\section{Conclusions and Outlook}

In this paper, we have investigated the dynamics of oblate droplets ($\mathrm{R/P} <1$), for which the electrohydrodynamic flow induces migration with velocity $U_{EHD}\sim1/h^2$ away from the wall, considerably stronger than the dielectrophoretic attraction to the wall ($U_{DEP}\sim1/h^4$), leading to a pronounced lift. This work highlights the importance of boundaries on droplet behavior in electric fields and develops a simple theory describing the migration velocity in terms of the leading flow singularity, a stresslet.  Since boundaries are always present in applications like microfluidics, droplet migration driven by EHD flow may play an important role. Our paper quantifies this effect and shows that the drift can be accurately predicted using asymptotic theory, which estimates droplet migration due to the stresslet-image-induced flow—a result that was not obvious a priori.

%The study can be extended to other fluid systems, e.g., for a prolate system ($\mathrm{R/P} >1$) the EHD and DEP effects are antagonistic, which may result in a steady hovering state, where the two effects balance. Fig.~\ref{fig:U_Prolate_hovering} illustrates this possibility based on theory and numerical simulations, and we hope to explore it experimentally in the future. In this particular system, the short-range repulsive DEP and long-range attractive EHD balance at an elevation of 3.5 drop radii above the wall.  We hope this prediction will stimulate further research into this phenomenon.  

%A leaky dielectric drop subject to an electric field perpendicular to a nearby wall will migrate away from the wall. This phenomenon arises due to the asymmetry introduced by the wall, and is reminiscent of the development of a dielectrophoretic force due to the asymmetry in the electric field at the drop surface introduced by the electric field on the surface of a particle. Indeed, both forces are present in a leaky dielectric system. This electrohydrodynamic forcing ($U\sim1/h^2$) is considerably stronger than dielectrophoresis ($U\sim1/h^4$), though it is only applicable in leaky dielectric systems with $\Rr/\Sr <1$.

%This work has described a means to manipulate soft matter and develops a simple theory describing the migration velocity in terms of the flow singularity, a stresslet,

\section*{Acknowledgments}
The authors would like to thank Dr. Paul Salipante for generously sharing his expertise, and Matthew Brucks, Xi Wan, Lucas Pham and Dr. Jeffrey J. Richards for their help with measuring the fluid viscosities and permittivities.
This research was supported by NSF award CBET-2126498 (PV and DS).
\appendix

%{\color{blue}

\section{Stokes Green's functions for the flow near a plane wall}
\label{sec:appendix:Stokes_Geens_wall}

In their seminal work, Blake and Chwang derived the Stokes Green's function associated with a point force near a plane wall \cite{blake1971note,blake_chwang1974viscous_flow_singularities, pozrikidis1992BIM_book}. The velocity tensor near a wall located at $x=x_w$ is given by: 
\begin{equation}
    \begin{aligned}
        G^w_{ij}(\bm{x}_0;\bm{x})= \,G^{FS}_{ij}(\bm{r})- G^{FS}_{ij}(\tilde{\bm{r}})
        + 2\,d^2_0\, D_{ij}(\tilde{\bm{r}})-2\,d_0\, Q_{ij}(\tilde{\bm{r}}),
        \label{eq:BEM:Greens_stokes_wall}
    \end{aligned}
\end{equation}
where $d_0 = x_0-x_w$ is the distance of the point force from the wall. Here, $\bm{\tilde{x}}_0$ denotes the mirror image of $\bm{x}_0$ with respect to the wall, with  $\bm{r} = \bm{x}_0-\bm{x}, ~r=|\bm{r}|$ and $\bm{\tilde{r}} = \bm{\tilde{x}}_0-\bm{x}, ~\tilde{r}=|\bm{\tilde{r}}|$. The first two terms on the right-hand side of \eqref{eq:BEM:Greens_stokes_wall} represent the primary and image Stokeslets in free space. The terms $D_{ij}$ and $Q_{ij}$ correspond to the potential dipole and point force doublet, respectively:
    \begin{align}
        G^{FS}_{ij}(\bm{r})&=\dfrac{\delta_{ij}}{r}+\dfrac{r_i \, r_j}{r^3},\label{eq:BEM:G_FS} \\
        D_{ij}(\bm{r})&=\pm \bigg(\dfrac{\delta_{ij}}{r^3}-3\dfrac{r_i\, r_j}{r^5}\bigg),\label{eq:BEM:D_ij} \\
        Q_{ij}(\bm{r})&=r_1 D_{ij}(\bm{r}) \pm \bigg(\dfrac{\delta_{j1}\,r_i-\delta_{i1}\,r_j}{r^3}\bigg).
        \label{eq:BEM:Q_ij}
    \end{align}
The minus sign corresponds to $j=1$ ($x$ direction), and the positive sign applies to $j=2,3$ ($y$ and $z$ directions). 

The associated stress tensor for this Green's function is:
\begin{equation}
    \begin{aligned}
        T^w_{ijk}(\bm{x}_0;\bm{x})= T^{FS}_{ijk}(\bm{r})- T^{FS}_{ijk}(\tilde{\bm{r}})
        + 2\,d^2_0\, T^D_{ijk}(\tilde{\bm{r}})-2\,d_0\, T^Q_{ijk}(\tilde{\bm{r}}),
        \label{eq:BEM:Greens_stokes_tensor_wall}
    \end{aligned}
\end{equation}
where
\begin{align}
    &T^{FS}_{ijk}(\bm{r})=-6\dfrac{r_i \, r_j\, r_k}{r^5}, \label{eq:BEM:TFS}\\ 
    &T^D_{ij}(\bm{r})=\pm 6\bigg(\!\!-\dfrac{\delta_{ij}r_k + \delta_{ik}r_j + \delta_{kj}r_i}{r^5}+5\dfrac{r_i\, r_j\, r_k}{r^7}\bigg), \label{eq:BEM:TD}\\ 
    &T^Q_{ij}(\bm{r})=r_1 T^D_{ijk}(\bm{r}) \pm 6\bigg(\dfrac{\delta_{ik}\,r_j\,r_1 - \delta_{j1}\,r_i\, r_k}{r^5}\bigg). \label{eq:BEM:TQ}
\end{align}
Similarly, the minus sign corresponds to $j=1$, and the positive sign applies to $j=2,3$.

\bibliography{paper_final_draft_arxiv}
\end{document}